%% file: main.tex
\documentclass[letterpaper, 10 pt, conference]{ieeeconf}  

\IEEEoverridecommandlockouts                              

\usepackage[utf8]{inputenc}
\usepackage[english]{babel}
\usepackage{graphicx}
\usepackage{float}
\usepackage{amssymb}
\usepackage{amsmath}
\usepackage{mathtools}
\usepackage{xcolor}
\usepackage{xspace}
\usepackage{algorithm}
\usepackage{multicol}
\usepackage{varwidth}
\usepackage{wrapfig}
\usepackage[noend]{algpseudocode}
\usepackage{marginnote}
\usepackage{multicol}
\usepackage{comment}
\usepackage{hyperref}
\hypersetup{
	pdftitle={LQR-MAML},
	pdfauthor={Negin Musavi},
	colorlinks=true,
	citecolor={blue},
	linkcolor = {blue},
}

\newtheorem{assumption}{Assumption}
\newtheorem{definition}{Definition}
\newtheorem{lemma}{Lemma}
\newtheorem{theorem}{Theorem}

\newtheorem{condition}{Condition}

\title{\LARGE \bf
Convergence of Gradient-based MAML in LQR
}

\author{Negin Musavi$^{1}$, and Geir E. Dullerud$^{2}$
\thanks{$^{1}$N. Musavi is with the Department of Mechanical Science
and Engineering, University of Illinois at Urbana-Champaign
        {\tt\small nmusavi2@illinois.edu}}%
\thanks{$^{2}$G. E. Dullerud is with the Department of Mechanical Science
and Engineering, University of Illinois at Urbana-Champaign
        {\tt\small dullerud@illinois.edu}}%
}

\begin{document}

\maketitle
\thispagestyle{empty}
\pagestyle{empty}

\input{abstract.tex}
\input{introduction.tex}
\input{problem.tex}
\input{lqr_maml.tex}
\input{simulation.tex}

\input{conclusion.tex}

\bibliographystyle{IEEEtran}
\bibliography{main.bib}

\input{appendixA.tex}
\input{appendixB.tex}

\input{appendixC.tex}
\input{appendixD.tex}

\end{document}

%% file: abstract.tex
\section{Abstract}\label{sec:abstract}

The main objective of this research paper is to investigate the local convergence characteristics of Model-agnostic Meta-learning (MAML) when applied to linear system quadratic optimal control (LQR). MAML and its variations have become popular techniques for quickly adapting to new tasks by leveraging previous learning knowledge in areas like regression, classification, and reinforcement learning. However, its theoretical guarantees remain unknown due to non-convexity and its structure, making it even more challenging to ensure stability in the dynamic system setting. This study focuses on exploring MAML in the LQR setting, providing its local convergence guarantees while maintaining the stability of the dynamical system. The paper also presents simple numerical results to demonstrate the convergence properties of MAML in LQR tasks.

%% file: introduction.tex
\section{Introduction}\label{sec:intro}
In the field of machine learning, the ability to quickly learn a new task with limited data by utilizing prior learning experiences is highly desirable. This approach is known as meta-learning or learning from learning. MAML is a well-known approach that can train machine learning models to swiftly adapt to new tasks with only a small amount of task-specific data. The concept behind MAML is to train a model on a group of related tasks so that it can acquire a useful starting parameter for a new task. MAML has gained recognition in scenarios such as few-shot image classification and reinforcement learning.

This paper focuses on applying MAML to a specific class of linear system quadratic optimal controllers, known as LQR. The goal is to examine the convergence of the algorithm for a collection of LQR tasks that vary in their system parameters or cost parameters.

There is a substantial body of research demonstrating the success of MAML through empirical studies on regression, classification, and reinforcement learning~\cite{al2017continuous, behl2019alpha, grant2018recasting, nichol2018first}. However, there are only a few works that have analyzed the theoretical guarantees of this algorithm. To give a few examples, the work in~\cite{fallah2020convergence, JM:JNSA2011, abbas2022sharp} have established local convergence guarantees for MAML in supervised learning such as classification and regression. The works in~\cite{collins2022maml, wang2022global} have explored its global convergence in specific settings. In reinforcement learning, the local convergence properties of the algorithm have been studied in~\cite{fallah2021convergence} under the assumption of having access to biased stochastic gradients and that stability of the dynamical system at each iteration of the algorithm. Another relevant study is the examination of the optimization landscapes of MAML in the LQR setting~\cite{molybog2021does}. This work provides conditions for stability and global convergence of the algorithm in the single task LQR setting, but doesn't include conditions for the multi-task setting. 

This paper presents the first examination of the convergence of the MAML algorithm for the multi-task LQR problem while also presenting conditions to guarantee the stability of the dynamical system. The MAML objective function for LQR tasks is not convex and does not inherit the gradient dominance property of the LQR cost function, which would have been helpful in analyzing convergence and ensuring the global convergence of gradient-based algorithms for the LQR problem. The study focuses on the local convergence properties of the algorithm and provides conditions to ensure the stability of the dynamical system. Whether global convergence is possible remains an outstanding research problem.

%% file: problem.tex
\section{Preliminaries and Problem Statement}\label{sec:problem}
In this section, we present the basics of the standard LQR problem followed by a brief overview of the gradient-based MAML method and its modification for LQR tasks. 

\subsection{Notation}\label{sec:notations}

We use the following mathematical notation throughout the paper. The set of real numbers is denoted by $\mathbb{R}$. For a real matrix $Z$, $Z^T$ represents its transpose, $\|Z\|$ its maximum singular value, $\|Z\|_{F}$ its Frobenius norm, $\hbox{tr}(Z)$ its trace, $\sigma_{\min}(Z)$ its minimum singular value, and $\hbox{vec}(Z)$ its vectorization obtained by stacking its columns. For a real square matrix $Z$, its spectral radius is denoted by $\hbox{rad}(Z)$. For a real symmetric matrix $Z$, $Z \succ 0$ and $Z \succeq 0$ indicate that $Z$ is positive definite and positive semi-definite, respectively. The open ball of radius $r > 0$ centered at $Z_{0} \in \mathbb{R}^{n \times m}$, is defined as $\mathcal{B}(Z_{0}, r) = \{Z \in \mathbb{R}^{n \times m} : \|Z- Z_{0}\| < r\}$. For matrices $Z_{1}$, $Z_{2}$, $\langle Z_1, Z_2 \rangle = \hbox{tr}(Z_{1}^{T}Z_2)$ denotes their inner product and $Z_1 \otimes Z_2$ denotes their Kronecker product. 

\subsection{Standard LQR Problem}\label{sec:lqr-prelim}
Consider the infinite horizon discrete-time LQR problem,
\begin{equation}\label{eq:lqr}
    \begin{aligned}
        &\hbox{minimize}_{u(.)}\ \ \mathbb{E}\bigg[\sum_{t=0}^{\infty} (x_t^TQx_t + u^{T}_{t} R u_{t})\bigg]\\
        &\hbox{subject to}\ \ x_{t+1} = Ax_t + Bu_t,\ \ x_0 \sim \mathcal{D},
    \end{aligned}
\end{equation}
where the initial state $x_0$ is randomly drawn from a distribution $\mathcal{D}$. The matrices $A \in \mathbb{R}^{n \times n}$ and $B \in \mathbb{R}^{n \times m}$ represent the system dynamics, and $Q \in \mathbb{R}^{n \times n}$ and $R \in \mathbb{R}^{m \times m}$ parameterize the cost, with $Q$ and $R$ being positive definite matrices. For a control policy at time $t \geq 0$ that is parameterized by a matrix $W \in \mathbb{R}^{m \times n}$ given by $u_t = -Wx_t$ the cost can be expressed as:
\begin{equation}\label{eq:lqr-cost}
    \begin{aligned}
        C(W) = \mathbb{E}_{x_0 \sim \mathcal{D}}\bigg[\sum_{t=0}^{\infty} x_t^T(Q + W^TRW)x_t\bigg].
    \end{aligned}
\end{equation}
This problem has an optimal solution as:
\begin{align*}\label{eq:opt-w}
    W^{*}_{lqr} = (R + B^TPB)^{-1}B^TPA,
\end{align*}
where $P \succ 0$ satisfies the Algebraic Riccati Equation $P = Q + A^TPA + A^TPB(R + B^TPB)^{-1}B^TPA$. 

Clearly $W$ is an stabilizing parameter for the LQR problem in~(\ref{eq:lqr}) with system parameters $(A, B)$ if $\hbox{rad}(A-BW) < 1$. Now we present explicit formulas for LQR cost, its gradient and its Hessian at an stabilizing $W$ which will be used later in our analysis.
\begin{itemize}
\item \textbf{LQR cost}: We can express the LQR cost as $$C(W) = \mathbb{E}_{x_0 \sim \mathcal{D}} \big[x_0^T P_W x_0\big],$$ with $P_W$ satisfying the following Lyapunov equation 
\begin{equation}\label{eq:ric}
    P_W = Q + W^TRW + (A-BW)^TP_W(A-BW).
\end{equation}
\item \textbf{LQR cost gradient}: We can express the gradient of the LQR cost as $$\triangledown C(W) = 2 E_W \Sigma_W,$$ with $E_W = (R + B^TP_WB)W - B^TP_WA$ and $\Sigma_W = \mathbb{E}_{x_0 \sim \mathcal{D}} \big[\sum_{t=0}^{\infty} x_tx_t^T\big]$ the state correlation matrix. If we further assume that $\mathbb{E}_{x_0 \sim \mathcal{D}} \big[x_0x_0^T\big]$ is full rank, then the unique solution for $\triangledown C(W) = 0$ is $W^{*}_{lqr}$(for more details see~\cite{fazel2018global}).
\item \textbf{LQR cost Hessian}: The action of Hessian operator on $Y \in \mathbb{R}^{n \times m}$ can be expressed as~\cite{bu2019lqr}:
\begin{equation}\label{eq:lqr-hessian}
    \begin{aligned}
        \triangledown^2 C(W)[Y,Y] &= 2 \langle(R + B^T P_W B) Y \Sigma_{W}, Y \rangle\\
        &\ \ \ - 4 \langle B^T P'_{W}[Y] (A-BW) \Sigma_{W}, Y \rangle,
    \end{aligned}    
\end{equation}
with $P'_{W}[Y] \succ 0$ satisfying the following equation $P'_{W}[Y] = (A-BW)^T P'_{W}[Y] (A-BW) + Y^{T} E_{W} + E_{W}^TY$.
\end{itemize}

\subsection{MAML}
MAML was first introduced in~\cite{finn2017model}. It is a meta-learning approach that aims to learn a good initialization for a model, such that it can quickly adapt to new tasks. More formally, suppose we have a set of tasks $\mathcal{I} = \{\mathcal{I}_1, \cdots \mathcal{I}_{K}\}$ drawn from distribution $p(\mathcal{I})$. Let the objective of task $\mathcal{I}_{i}$ be as minimizing a loss function that is parameterized by parameter $W$, i.e. $\min_{W} \mathcal{L}_{i}(W)$.

The idea behind MAML, under the assumption that the task can be solved by gradient descent, is to fine-tune parameter $W$ for set of tasks $\mathcal{I}$ such that one or a few gradient steps can be taken with respect to a particular task $\mathcal{I}_i$ allowing it to serve as a good initialization for a new task drawn from the same distribution as the other tasks. For instance, with one gradient step this is achieved by solving the following optimization problem:
\begin{align*}
    \min_{W} \sum_{\mathcal{I}_i \sim p(\mathcal{I})} \mathcal{L}_{i}(W - \eta \triangledown \mathcal{L}_{i}(W)),
\end{align*}
with $\eta > 0$ as a step-size parameter.

\subsection{Problem Statement}
We devote this section to introducing the Gradient-based MAML for set of LQR tasks. Let us first outline the LQR task set $\mathcal{I}$.

\textbf{\begin{definition}\label{def:task}
\textnormal{Each LQR task $\mathcal{I}_{i}$ is defined by a tuple of matrices $\big(A_{i},B_{i}, Q_{i}, R_{i}\big)$, with $A_{i} \in \mathbb{R}^{n \times n}$, $B_{i} \in \mathbb{R}^{n \times m}$, and $Q_{i}  \in \mathbb{R}^{n \times n}$, $R_{i}  \in \mathbb{R}^{m \times m}$ being positive definite matrices. The cost function associated with each task, $C_{i}(W):\mathbb{R}^{m \times n} \rightarrow \mathbb{R}^{\geq 0}$, is parameterized by $W \in \mathbb{R}^{m \times n}$ as in (\ref{eq:lqr-cost})}.
\end{definition}}
\vspace{0.25cm}
The objective is to find the best $W$ for the tasks $\mathcal{I}$ using MAML approach such that it can be a good starting point for a new LQR task. This amounts to the following MAML objective function:
\begin{equation}\label{eq:maml-cost}
\begin{aligned}
    F(W) &:= \sum_{\mathcal{I}_i \sim p(\mathcal{I})} F_{i}(W)
\end{aligned}
\end{equation}
with $F_{i}(W) =  C_{i}(W - \eta \triangledown C_{i}(W))$ and $\eta >0$ as the step-size parameter. The MAML objective is to be minimized using gradient descent, with a single step update rule as follows \begin{align}\label{eq:gradient_step}
W \leftarrow W - \beta \triangledown F(W),
\end{align}
with step-size parameter $\beta > 0$. This is called gradient-based MAML and is depicted in Algorithm~\ref{Al:MAML}. The step-sizes $\eta$ and $\beta$ are also known as the inner-loop and outer-loop step-sizes. Define $G_{i}(W) := \big(I - \eta \triangledown^2 C_{i}(W)\big)$, then we have:
\begin{equation}\label{eq:grad-F}
\begin{aligned}
    \triangledown F(W) &= \sum_{\mathcal{I}_i \sim p(\mathcal{I})} \triangledown F_{i}(W)\\
    &= \sum_{\mathcal{I}_i \sim p(\mathcal{I})} G_{i}(W) \triangledown C_{i}(W - \eta \triangledown C_{i}(W)).
\end{aligned}
\end{equation}

\begin{algorithm}[ht!]
    \caption{\small gradient-based MAML for LQR tasks.}\label{Al:MAML}
    \begin{algorithmic}[1]
        \State Initialize $W$
	    \While{$not\ done$} 
            \State Choose step-size parameters $\eta$ and $\beta$
    	    \ForAll{$tasks\ \in \mathcal{I}$} 
    	        \State Evaluate $\triangledown C_{i}(W)$
                \State
                \begin{varwidth}[t]{\linewidth}
                Compute adapted parameter with a gradient
                \par
                step: $\tilde{W}_{i} = W - \eta \triangledown C_{i}(W)$
                \end{varwidth}
                \State Evaluate $\triangledown C_{i}(\tilde{W}_{i})$ and $G_{i}(W)$
    	    \EndFor
            \State Update $W \leftarrow W - \beta \sum_{\mathcal{I}_i \sim p(\mathcal{I})} G_{i}(W) \triangledown C_{i}(\tilde{W}_{i})$
	    \EndWhile
     \Return Estimation of MAML optimal parameter $\hat{W}^{*}_{maml}$
    \end{algorithmic}
\end{algorithm}

%% file: lqr_maml.tex
\section{Convergence Analysis of Gradient-based MAML for Multi-task LQR Problem}\label{sec:lqr-maml-multi-task}

In this section, we will examine the local convergence of gradient-based MAML for a set of LQR tasks.  We will begin by presenting definitions, assumptions, and lemmas, and then employ them to prove the desired result. For the sake of conciseness, only some of the results will be accompanied by their proofs. We commence with the following definition.

\textbf{\begin{definition}
\textnormal{We say:
\begin{itemize}
\item parameter $W$ is task-stabilizing for task $\mathcal{I}_i$ if $\hbox{rad}(A_i-B_iW) < 1$; and
\item parameter $W$ is MAML-stabilizing if it is task-stabilizing for all tasks in $\mathcal{I}$ and $\hbox{rad}(A - B(W - \eta \triangledown C_{i}(W))) < 1$ holds for all tasks.
\end{itemize}}
\end{definition}}
\vspace{0.25cm}

Suppose parameter $W$ is task-stabilizing for all tasks. Define $\mu = \sigma_{min}\bigg(\mathbb{E}_{x_0 \sim \mathcal{D}}[x_0x_0^T]\bigg)$, and $\delta_{i}(W)$ as $$\delta_{i}(W) = \dfrac{\sigma_{\min}(Q_{i})\mu}{4C_{i}(W)\|B_i\|(\|A_i-B_iW\| +1)}.$$ Also let $\delta(W) = \min_{\mathcal{I}_i} \delta_{i} (W)$. Given these, the following lemma provides a sufficient condition on $\eta$ to ensure $W$ be also MAML-stabilizing.

\textbf{\begin{lemma}\label{lem:suff-stable}
\textnormal{Given that $W$ is task-stabilizing for all tasks, if the step-size $\eta$ satisfies the condition $\eta<\min_{\mathcal{I}_i}\bigg\{\dfrac{\delta_{i}(W)}{\|\triangledown C_{i}(W)\|}\bigg\},$ then it is also MAML-stabilizing.}
\end{lemma}}
\vspace{0.25cm}

This outcome is a direct consequence of the results in~\cite{fazel2018global} to guarantee that for each task the control policies $W-\eta \triangledown C_{i}(W)$ stabilizes the system. Before presenting the convergence theorem for Algorithm~\ref{Al:MAML}, we need to make certain assumptions regarding each task's cost function.

\textbf{\begin{lemma}\label{ass:LQR-cost-assumptions}
\textnormal{Given that $W$ and $U$ are task-stabilizing for task $\mathcal{I}_i$, there exists $\theta_{i}(W)$, $\ell_{i}(W)$, $\rho_{i}(W)$, and $H_{i}(W)$ such that:
\begin{align*}
    &|C_{i}(U) - C_{i}(W)| \leq \mathbb{E}_{x_0\sim \mathcal{D}}\big[\|x_0\|^2\big] \theta_{i}(W)  \|U-W\|,\\
    &\|\triangledown C_{i}(U) - \triangledown C_{i}(W)\| \leq \ell_{i}(W)  \|U-W\|,\\
    &\|\triangledown^{2} C_{i}(U) - \triangledown^{2} C_{i}(W)\| \leq  \rho_{i}(W)  \|U-W\|,\\
    &\hbox{vec}\big(\triangledown C_{i}(U) - \triangledown C_{i}(W)\big) =  H_{i}(W)  \hbox{vec}\big(U-W\big).
\end{align*}}
\end{lemma}}
\vspace{0.25cm}
The first three inequalities indicate that the LQR cost function, its gradient, and its Hessian are locally Lipschitz continuous at any stabilizing $W$. Hence an upper bound for $\theta_{i}(W)$ and $\ell_{i}(W)$ are provided in references ~\cite{fazel2018global, jansch2020policy} in terms of $W$, system parameters and cost parameters. We have also obtained an upper bound for $\rho_{i}(W)$, as presented in Lemma~\ref{lem:pert-hess-cost}, with a brief proof that is available in Appendix \textcolor{blue}{B} for better understanding. The final statement follows directly from the mean-value theorem applied to the LQR gradient, as the LQR gradient is locally Lipschitz continuous at any stabilizing $W$. Lemma~\ref{lem:mean-value-grad-cost} provides the details of this and its proof is also included in Appendix \textcolor{blue}{B}. The subsequent assumption pertains to the set of LQR problems that will be considered.

\textbf{\begin{assumption}\label{ass:tasks}
\textnormal{The set of LQR tasks $\mathcal{I}$ share the same system matrices, namely $A_{i}=A$ and $B_{i}=B$ for all $\mathcal{I}_i$. The cost parameters $Q_{i}$ and $R_{i}$ vary among the tasks.}
\end{assumption}}
\vspace{0.25cm}

Furthermore, similar to the analysis presented in~\cite{fallah2020convergence}, it is required that the variance of the LQR cost gradient be bounded over the set of tasks $\mathcal{I}$. This requirement is further elaborated in the following assumption. The set of all parameters that stabilize task $\mathcal{I}_i$ is denoted as $\mathcal{S}_i =\{W: \hbox{rad}(A_i-B_iW) < 1\}$. Also, let $\mathcal{S} = \cap_{i} \mathcal{S}_i$. Under Assumption~\ref{ass:tasks}, the sets $\mathcal{S}_i$ are identical for all tasks, and therefore $\mathcal{S}$ is non-empty and identical to each $\mathcal{S}_i$.

\textbf{\begin{assumption}\label{ass:variance}
\textnormal{Let $\mathcal{W}$ be a compact subset of $\mathcal{S}$. There exists a constant $\sigma > 0$ such that for all $W \in \mathcal{W}$ $$\mathbb{E}_{i}\bigg[\|\triangledown C_i(W) - \mathbb{E}_{i}  \big[\triangledown C_i(W)\big]\|_F^{2}\bigg] \leq \sigma^{2}.$$} 
\end{assumption}}
\vspace{0.25cm}

With these assumptions in place, we can proceed to the next two lemmas. The following lemma provides a sufficient condition on $\beta$ to ensure that $W-\beta \triangledown F(W)$ is stabilizing for all tasks given that $W$ is stabilizing for all tasks. 

\textbf{\begin{lemma}\label{lem:beta}
\textnormal{Let $W$ be an stabilizing parameter for all tasks in Assumption~\ref{ass:tasks}. Define $\bar{\delta}_{i}(W)$ as:
\begin{align*}
     \bar{\delta}_{i}(W) = \dfrac{\sqrt{\|A-BW\|^{2} + \frac{\mu\sigma_{min}(Q_{i})}{C_{i}(W)}} - \|A-BW\|}{\|B\|}.
\end{align*}
If $\beta$ satisfies the following condition
\begin{align*}
     \beta < \min_{\mathcal{I}_{i}} \bigg\{ \dfrac{\bar{\delta}_{i}(W)}{\|\triangledown F(W)\|} \bigg\},
\end{align*}
then $W- \beta \triangledown F(W)$ is stabilizing for all tasks in $\mathcal{I}$.} 
\end{lemma}}
\vspace{0.25cm}

The proof of this lemma is deferred to Appendix \textcolor{blue}{C}. Now let $\ell(W) = \max_{i} \ell_{i}(W)$, $\bar{\ell}(W) = \max_{i} \ell_{i}(W - \eta \triangledown C_{i}(W))$, $\|H(W)\| = \max_{i} \|H_{i}(W)\|$, and $\rho(W) = \max_{i} \rho_{i}(W)$ for a given $W$. Then we have:

\textbf{\begin{lemma}\label{lem:pert-maml-grad-cost}
\textnormal{Consider the MAML objective in~(\ref{eq:maml-cost}). Suppose $W$ and $U$ are task-stabilizing for all tasks in Assumption~\ref{ass:tasks}. If the learning rate $\eta > 0$ is such that $\eta < \min_{\mathcal{I}_i}\bigg\{\dfrac{\delta(W)}{\|\triangledown C_{i}(W)\|}, \dfrac{\delta(U)}{\|\triangledown C_{i}(U)\|}\bigg\},$ then we have:
\begin{align*}
    \|\triangledown F(U) - \triangledown F(W)\| \leq L(W) \|U-W\|,
\end{align*}
with $L(W) = \bar{\ell}(W)\big(1 + \eta (\ell(W) + \delta(W)\rho(W))\big)(1+\eta \ell(W))+\eta \rho(W) (1 + \eta \|H(W)\|)\| \mathbb{E}_{i} \big[\|\triangledown C_{i}(W)\|_{F}\big]$.}
\end{lemma}}
\vspace{0.25cm}

 The proof of this lemma is derived from the definition of the MAML objective, by applying Lemmas~\ref{lem:suff-stable} and \ref{ass:LQR-cost-assumptions}, by utilizing Assumption~\ref{ass:variance}, and follows a similar approach to ~\cite{fallah2020convergence}. The proof of this lemma is deferred to Appendix \textcolor{blue}{C}. Also consider the following conditions for both step-sizes.
 
 \textbf{\begin{condition}\label{ass:stepsize}
\textnormal{For a task-stabilizing $W_{j}$ of all tasks at iteration $j$ of Algorithm~\ref{Al:MAML}, consider the following conditions on $\eta$:
\begin{align*}
    \eta_{j} &< \min_{\mathcal{I}_i} \bigg\{ \dfrac{1}{ 4\|\triangledown^{2} C_{i}(W_{j})\| + 2\rho_{i}(W_j)\delta_{i}(W_j) },\\
    &\ \ \ \ \ \ \ \ \ \ \ \dfrac{(1 - \alpha)\delta(W_{j})}{\alpha \delta(W_{j}) \ell_{i}(W_{j}) + \|\triangledown C_{i}(W_{j})\|}\bigg\},
\end{align*}
for some $\alpha \in (0, 1)$, and then on $\beta$:
\begin{align*}
    \dfrac{a}{L(W_{j})} < \beta_j < \dfrac{b}{L(W_{j})} < \min_{\mathcal{I}_i} \bigg\{\dfrac{\alpha \bar{\delta}_{i}(W_{j})}{\|\triangledown F(W_{j})\|}\bigg\},
\end{align*}
for some $a,b$ such that $2a > b^2$.}
\end{condition}}
\vspace{0.25cm}

The stability of the linear system during the optimization and the convergence of the algorithm, as stated in the following theorem, depend on satisfying these conditions for $\eta$ and $\beta$. Specifically, the last conditions on $\eta$ and $\beta$ serve the former purpose, while the first condition on $\eta$ and the first two conditions on $\beta$ are for the latter.

\textbf{\begin{theorem}\label{th:maml-convergence-stationary}
\textnormal{Let $\mathcal{W}$ be a compact subset of $\mathcal{S}$. Initialize the Algorithm~\ref{Al:MAML} with a task-stabilizing $W_0$ for all the tasks in Assumption~\ref{ass:tasks} such that $W_0 \in \mathcal{W}$. Let $W^{*}_{maml} := \arg \min_{W} F(W)$, and define the sub-optimality gap $\Delta = F(W_{0}) - F(W^{*}_{maml})$. Given that Assumption~\ref{ass:variance} holds, if the step-sizes $\eta$ and $\beta$ satisfy the conditions in Condition~\ref{ass:stepsize} at each iteration, then for any desired accuracy $\epsilon > 0$, there exist constants $\tilde{c}_{1}, \tilde{c}_2 >0$ such that after running Algorithm~\ref{Al:MAML} for at most $\dfrac{\Delta\big(\tilde{c}_{1} + \tilde{c}_{2} \big(\sigma + \epsilon\big)\big)}{(2a - b^{2})\epsilon^{2}}$ iterations, it will find a solution $W_{\epsilon}$ that satisfies $$\|\triangledown F(W_{\epsilon})\| \leq \epsilon,$$ where $\sigma$ is the variance constant from Assumption~\ref{ass:variance}, and $a, b$ are constants defined in Condition~\ref{ass:stepsize}.}
\end{theorem}}
\vspace{0.25cm}

The theorem states that with proper choices of step-sizes, MAML will reach a stationary point, characterized by a small gradient magnitude ($\leq \epsilon$), in a number of iterations that grows inversely proportional to the square of $\epsilon$. In other words, the computational cost of finding the stationary point scales as $\mathcal{O}(1/ \epsilon^{2})$. The proof of this theorem is deferred to Appendix \textcolor{blue}{D} to enhance readability. 

%% file: simulation.tex
\section{Simulation Results}\label{sec:sim}

In this part we provide simulation results for the convergence of the gradient descent-based MAML algorithm. Specifically, we tested the algorithm on ten LQR tasks with identical $A$ and $B$ matrices, where $A = \begin{pmatrix} 1.5 & 0 \\ 0 & -2 \end{pmatrix}$ and $B = \begin{pmatrix} 0.5 \\ 0.5 \end{pmatrix}$. The LQR tasks are constructed with different cost parameters $Q_i$ and $R_i$, where $Q_i$ are a linear combination of $\bar{Q}_1 = \begin{pmatrix} 0.01 & -0.5 \\ -0.5 & 200 \end{pmatrix}$ and $\bar{Q}_2 = \begin{pmatrix} 200 & 1 \\ 1 & 0.01 \end{pmatrix}$, and $R_i$ is a linear combination of $\bar{R}_1 = 2$ and $\bar{R}_2 = 0.1$ as:
\begin{align*}
    Q_i &= \alpha_{1i}\bar{Q}_1 + \alpha_{2i}\bar{Q_2}\\
    R_i &= \alpha_{3i}\bar{R}_1 + \alpha_{4i}\bar{R_2},
\end{align*}
with coefficients $\alpha_{1i}$, $\alpha_{2i}$, $\alpha_{3i}$, and $\alpha_{4i}$ that are randomly drawn from the interval $(0, 10)$. 

\begin{figure}[!htpb]
    \centering
    \includegraphics[width=1\linewidth]{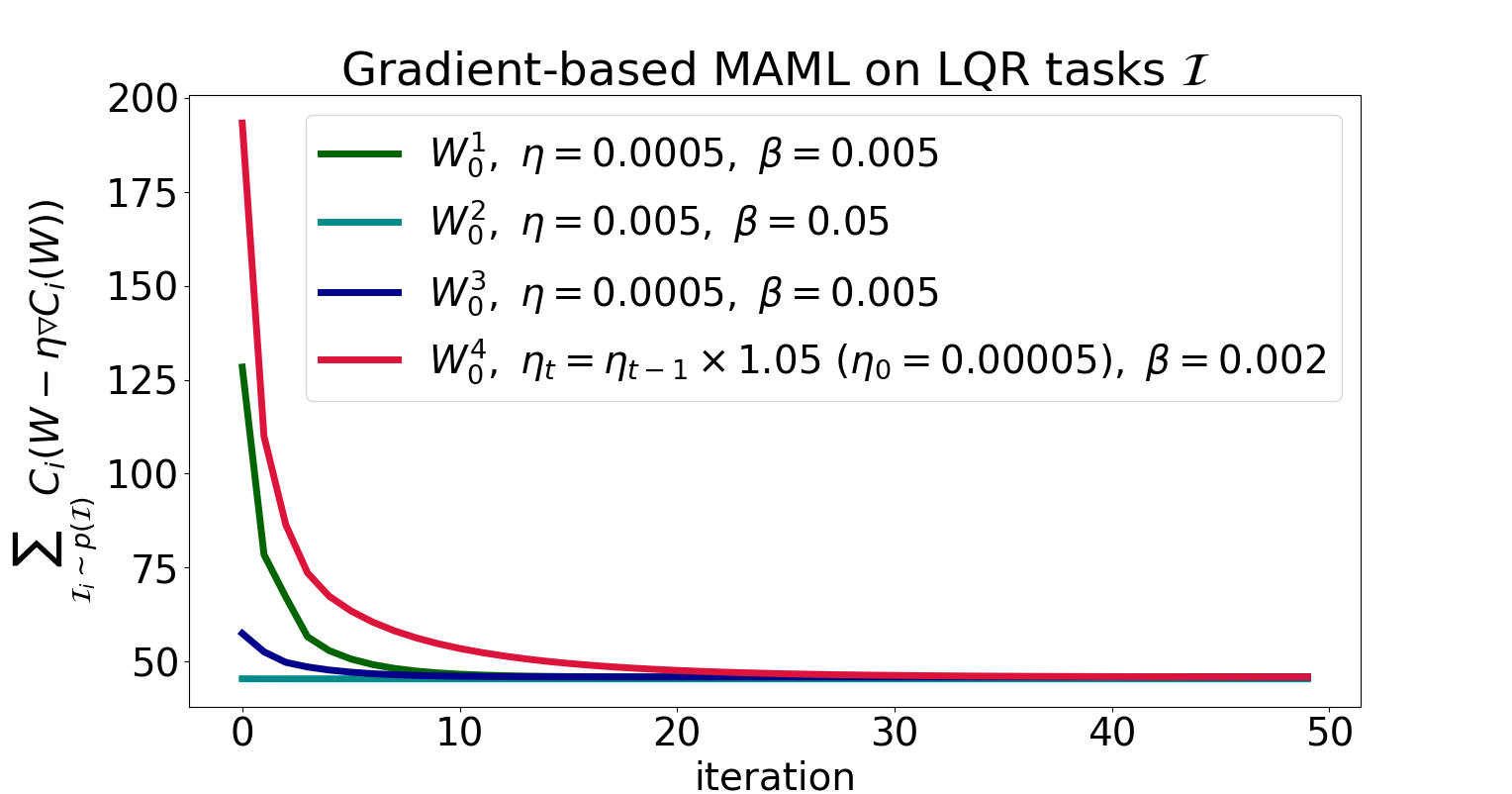}
    \caption{\small Convergence of Algorithm~\ref{Al:MAML} for tasks $\mathcal{I}$ with different initialization $W_{0}^{1}$,$W_{0}^{2}$, $W_{0}^{3}$, and $W_{0}^{4}$.}
    \label{fig:convergence}
\end{figure}

We run Algorithm~\ref{Al:MAML} with four different initialization and choices of step-sizes $\eta$, $\beta$. The results of the experiment, which ran for 50 iterations, are presented in Figure~\ref{fig:convergence}. In three of the runs, we kept the inner-loop step-size $\eta$ constant throughout the optimization process, while in one run, we varied it. In the plot where the initialization $W_{0}^{4}$ is far from the converging point, we started with a small $\eta$ ($=0.00005$) and gradually increased it. The reason for this is that in this case, $W_{0}^{4}$ is close to the boundary of the stabilizing set $\mathcal{S}$ and requires a lower $\eta$ to ensure task stability of all tasks. As $W_j$ moves away from the boundaries of the set $\mathcal{S}$, a higher $\eta$ can be chosen to speed up the convergence, so we increased the step-size accordingly. It is noted that in this case at iteration $50$ the step-size $\eta = 0.0066$.

The optimal parameter returned by the Algorithm~\ref{Al:MAML}, i.e. $\tilde{W}^{*}_{maml}$, for the four runs is recorded and is presented in Table~\ref{tab:convergence}. Here, $w_1$ and $w_2$ represent the first and second entries of $W$.
\begin{table}[!htpb]
\small 
\centering
\def\arraystretch{1.3}
\setlength\tabcolsep{3pt}
\caption{Output of the Algorithm~\ref{Al:MAML} for different runs.}\label{tab:convergence}
\begin{tabular}{|c|c|c|c|c|}
\hline
{\bf $W_{0}$} & $W_{0}^{1}$ & $W_{0}^{2}$ & $W_{0}^{3}$ & $W_{0}^{4}$\\
\hline
{\bf $\hat{W}^{*}_{maml}$\/}  & $\begin{matrix} w_1 = +1.38 \\ w_2 = -2.13\end{matrix}$ & $\begin{matrix} w_1 = +1.36 \\w_2 = -2.12\end{matrix}$ & $\begin{matrix} w_1 = +1.38 \\w_2 = -2.13\end{matrix}$ & $\begin{matrix} w_1 = +1.37 \\ w_2 = -2.13\end{matrix}$\\
\hline
\end{tabular}
\end{table}
To demonstrate the efficiency of Algorithm~\ref{Al:MAML}, it is necessary to provide a measure of how closely these estimated optimal parameters match the MAML objective in (\ref{eq:maml-cost}). Clearly, the MAML objective depends on the inner-loop step-size $\eta$. Therefore, we girded the set of stabilizing parameters $\mathcal{S}$ with a resolution of $0.05$ and computed the MAML objective on the grid points using different $\eta$ values. We then searched for the optimal parameter for each and recorded the results in Table~\ref{tab:maml-grid}. According to these results, the optimal parameter is located within a ball of radius $0.23$ centered at $[1.347, -2.102]$ which we call it MAML optimal neighborhood. On closer inspection of the output of Algorithm~\ref{Al:MAML} recorded in Table~\ref{tab:convergence}, we find that all the parameters lie within this optimal neighborhood. Therefore, we can conclude that Algorithm~\ref{Al:MAML} converges to a MAML optimal neighborhood estimated by girding the $\mathcal{S}$.

\begin{table}[!htpb]
\small 
\centering
\def\arraystretch{1.3}
\setlength\tabcolsep{3pt}
\caption{The MAML objective's optimal point calculated by searching a gird created over the set $\mathcal{S}$.}\label{tab:maml-grid}
\begin{tabular}{|c|c|c|c|}
\hline
{\bf $\eta$} & 0.0005 & 0.0050 & 0.0066\\
\hline
{\bf $W^{*}_{maml} $\/}  & $\begin{matrix} w_1 = +1.402 \\ w_2 = -2.102 \end{matrix}$ & $\begin{matrix} w_1 = +1.292 \\ w_2 =  -2.102 \end{matrix}$ & $\begin{matrix} w_1 = +1.347 \\ w_2 = -2.102 \end{matrix}$\\
\hline
\end{tabular}
\end{table}

%% file: conclusion.tex
\section{Conclusion and Future Work}\label{sec:conc}

We studied the convergence of MAML for a set of specific LQR tasks that share the same dynamics. We derived that with conditions on both step-sizes the linear system stability is guaranteed and MAML reach any desired accuracy level of $\epsilon \geq 0$, with a computational cost that scales as $\mathcal{O}(1/\epsilon^2)$. Our focus was on the model-based version of the LQR problem, where we have exact information about the cost, its gradient, and its Hessian for a specific policy. However, we would like to point out that these results could also be extended to model-free scenarios, where the cost gradient and Hessian must be estimated from data. We aim to explore these model-free extensions in our future work. 

\section{Acknowledgement}
The authors would like to thank Sanjay Shakkottai, Dawei Sun, and Sayan Mitra for the discussions that helped lead to this work.

%% file: appendixA.tex
\section*{Appendix A}\label{app:lqr-detail}

In this part we provide useful results that are helpful in our later analysis. The next lemma combines few of these results borrowed from~\cite{fazel2018global}.

\textbf{\begin{lemma}\label{lem:helper_lemmas}
\textnormal{Suppose $W$ and $U$ are stabilizing for the linear system in (\ref{eq:lqr}) and $\|U-W\| \leq \delta(W)$. Let $m(W), n(W), \psi(W),  \phi(W)$ be defined as:
\begin{align*}
    m(W) &= 4\bigg(\dfrac{C(W)}{\sigma_{\min}(Q)}\bigg)^2\dfrac{\|B\|(\|A-BW\| +1)}{\mu}\\
    n(W) &= 6 \Bigg(\bigg(\dfrac{C(W)}{\sigma_{\min}(Q)}\bigg)^{2} \|W\|^{2}\|R\|\|B\|(\|A-BW\|+1)\\
    &\ \ \ \ \ \ \ \ + \bigg(\dfrac{C(W)}{\sigma_{\min}(Q)}\bigg) \|W\|\|R\|  \Bigg)\\
    \psi(W) &= \|R\| + \|B\|^{2} \dfrac{C(W)}{\mu} \\
    &\ \ \ \ \ \ \ \ + \bigg(\|B\|\|A\| + \|B\|^{2}\|W\| \bigg)n(W)\\
    \phi(W) &= 2 \bigg( \|R\|\|W\| + \|B\| \|A-BW\| \dfrac{C(W)}{\mu} \bigg), 
\end{align*}
Then it holds that:
\begin{align*}
    \|\Sigma_{U} - \Sigma_{W}\| &\leq m(W) \|U-W\|\\
    \|P_{U} - P_{W}\| &\leq n(W) \|U-W\|\\
    \|E_{U} - E_{W}\| &\leq \psi(W) \|U-W\|\\
    \hbox{tr}\big(\Sigma_{W}\big) &\leq \dfrac{C(W)}{\sigma_{\min}(Q)}\\
    \|P_{W}\| &\leq \dfrac{C(W)}{\mu}\\
    \|E_{W}\| &\leq \phi(W).
\end{align*}}
\end{lemma}}
\vspace{0.25cm}

The following Lemma is borrowed from~\cite{jansch2020policy} that is also helpful in our later analysis.

\textbf{\begin{lemma}\label{lem:p-prime}
\textnormal{Let $Y \in \mathbb{R}^{n \times m}$, then for $P'_{W}[Y]$ we have:
\begin{align*}
    \|P'_{W}[Y]\| \leq  \dfrac{\zeta(W)C(W)}{\mu} \|Y\|,
\end{align*}
with $\zeta(W) = \sqrt{\dfrac{1}{\sigma_{\min}(Q)}\bigg(\|R\| + \dfrac{1 + \|B\|^2}{\mu}C(W)\bigg) - 1}$.}
\end{lemma}}

%% file: appendixB.tex
\section*{Appendix B}
The next lemma is about local Lipschitz continuity of Hessian of the LQR cost.

\textbf{\begin{lemma}\label{lem:pert-hess-cost}
\textnormal{Suppose $W$ and $U$ are stabilizing for the linear system in (\ref{eq:lqr}) and $\|U-W\| \leq \delta(W)$. Then it holds that:
\begin{align*}
    \|\triangledown^2 C (W) - \triangledown^2 C (U)\| \leq \rho(W) \|W - U\|,
\end{align*}
with $\rho(W)$ in terms of $\dfrac{C(W)}{\mu \sigma_{\min}(Q)}$, $\mathbb{E}\big[\|x_0\|^2\big]$, $\|A\|$, $\|B\|$, $\|R\|$, $\dfrac{1}{\sigma_{\min}(R)}$, $\mu$, $\hbox{tr}(R)$, $\|B\|_{F}$, and $\|A - BW\|$.}
\end{lemma}}
\vspace{0.25cm}

\begin{proof}
Let $A_W = A-BW$, then we can write following for $W$ and $U$:
\begin{align*}    \| \triangledown^2 C(U) &- \triangledown^2 C(W)\|\\
    &= \sup_{\|Y\|_{2} = 1} \| \triangledown^2 C(U)[Y] - \triangledown^2 C(W)[Y]\|\\
    &= \sup_{\|Y\|_{2} = 1} | \triangledown^2 C(U)[Y,Y] - \triangledown^2 C(W)[Y,Y] |\\
    & \leq 2 \sup_{\|Y\|_{2} = 1} |\langle(R + B^T P_{U} B) Y \Sigma_{U}, Y \rangle \\
    &\ \ \ \ \ \ \ \ \ \ \ \ \ \ \ - \langle(R + B^T P_{W} B) Y \Sigma_{W}, Y \rangle| \\
    & + 4 \sup_{\|Y\|_{2} = 1} | \langle B^T P'_{U}[Y] A_U \Sigma_{U}, Y \rangle \\
    &\ \ \ \ \ \ \ \ \ \ \ \ \ \ \ - \langle B^T P'_{W}[Y] A_W \Sigma_{W}, Y \rangle |.
\end{align*}
We treat the two terms separately. For brevity we skip the complete proof. We can bound the first term as:
\begin{align*}
    |\langle(R &+ B^T P_{U} B) Y \Sigma_{U}, Y \rangle - \langle(R + B^T P_{W} B) Y \Sigma_{W}, Y \rangle|\\
    &\leq \|Y\|^2 \|B\|^2 \|P_{W} - P_{U}\| \hbox{tr}\big(\Sigma_{W}\big)\\
    &+ \|Y\|^2 \|\Sigma_{W} - \Sigma_{U}\| \bigg(\hbox{tr}(R) + \|B\|_{F}\|P_{U}\|\bigg).
\end{align*}
By applying Lemmas~\ref{lem:helper_lemmas} we have:
\begin{align*}
    |\langle&(R + B^T P_{U} B) Y \Sigma_{U}, Y \rangle - \langle(R + B^T P_{W} B) Y \Sigma_{W}, Y \rangle|\\
    &\leq  \|Y\|^{2} \bigg(n(W)\|B\|^{2}\dfrac{C(W)}{\sigma_{\min}(Q)}\\
    &\ \ \ \ \ \ \ \ \ + d(W) m(W)\|B\|_{F}\dfrac{C(W)}{\mu}\\
    &\ \ \ \ \ \ \ \ \ + m(W) \hbox{tr}(R) \bigg)\|U - W\|\\
    &=: a(W) \|Y\|^{2}\|U-W\|,
\end{align*}
where $d(W) = 1 + \dfrac{\mathbb{E}_{x_0\sim \mathcal{D}}\big[\|x_0\|^2\big]\theta(W)}{\delta(W) C(W)}$. Now let's work on the second term:
\begin{align*}
    | \langle &B^T P'_{U}[Y] A_U \Sigma_{U}, Y \rangle - \langle B^T P'_{W}[Y] A_W \Sigma_{W}, Y \rangle |\\ 
    &\leq \|B\|\|Y\|\|P'_{W}[Y]\|\Sigma_{W}-\Sigma_{U}\|\hbox{tr}(A_W)\\
    &+ \|B\|\|Y\|\|P'_{W}[Y]\|A_W-A_U\|\hbox{tr}(\Sigma_{U})\\
    &+ \|B\| \|P'_{W}[Y] - P'_{U}[Y]\|\|Y\|\hbox{tr}(A_U\Sigma_{U}).
\end{align*}
First consider the following helper lemma on bounding the perturbation of $P'_{W}[Y]$.

\textbf{\begin{lemma}\label{lem:pert-p-prime}
\textnormal{For stabilizing $W$ and $U$ such that $\|U-W\| \leq \delta(W)$, we have:
\begin{align*}
    \|P'_{W}[Y] - P'_{U}[Y]\| \leq e(W) \|Y\|\|U-W\|,
\end{align*}
with
\begin{align*}
    e(W) &= 4\bigg(\dfrac{C(W)}{\mu \sigma_{\min}(Q)} \psi(W)\\
    &\ \ \ \ \ \ + 8\bigg(\dfrac{C(W)}{\mu \sigma_{\min}(Q)}\bigg)^{2} \|B\|(\|A_W\|+1)\phi(W)\bigg).
\end{align*}}
\end{lemma}}
\vspace{0.5cm}

We leave this lemma without proof for brevity. 
we can continue as:
\begin{align*}
    | \langle & B^T P'_{U}[Y] A_U \Sigma_{U}, Y \rangle - \langle B^T P'_{W}[Y] A_W \Sigma_{W}, Y \rangle |\\ 
        &\leq  \|B\|\|Y\|^{2}C(W)\bigg(\dfrac{\zeta(W) m(W)\hbox{tr}(A_W)}{\mu}\\
        &\ \ \ \ \ \ \ \ \ \ \ \ \ \ \ \ \ \ \ \ \ \ \ \ \ \ \ + \dfrac{\|B\|\zeta(W) d(W)C(W)}{\mu\sigma_{\min}(Q)}\\
        &\ \ \ \ \ \ \ \ \ \ \ \ \ \ \ \ \ \ \ \ \ \ \ \ \ \ \ + \dfrac{d(W)e(W)}{\sigma_{\min}(Q)}\bigg)\|U-W\|\\
        &=: b(W)\|Y\|^{2} \|U-W\|.
\end{align*}
Combining the two bounds concludes the proof as:
\begin{align*}
    \| \triangledown^2 C(U) &- \triangledown^2 C(W)\|\\
    &\leq \sup_{\|Y\|_{2} = 1} \big(2 a(W) +  4b(W)\big) \|Y\|^{2} \|U-W\|\\
    &=: \rho(W) \|U-W\|.
\end{align*}   
\end{proof}

The following lemma states the existence of $H(W)$ introduced in the Assumption~\ref{ass:LQR-cost-assumptions}.

\textbf{\begin{lemma}\label{lem:mean-value-grad-cost}
\textnormal{For stabilizing $W$ and $U$ such that $\|U-W\| \leq \delta(W)$, there exists a matrix $H(W)$ such that have:
\begin{align*}
    \hbox{vec}\big(\triangledown C(U) - \triangledown C(W)\big) = H(W) \hbox{vec}\big(U - W\big),
\end{align*}
where $H(W)$ includes second order partial derivative of LQR cost evaluated at parameters on the line connecting $\hbox{vec}(W)$ and $\hbox{vec}(U)$. Also $\|H(W)\| \leq \|\triangledown^2 C(W)\| + \rho(W)\delta(W)$.} 
\end{lemma}}
\vspace{0.25cm}

\begin{proof}
If conditions of the Assumption~\ref{ass:LQR-cost-assumptions} are met then for such $U$ we have:
\begin{align*}
    \|\triangledown^2 C(U) - \triangledown^2 C(W)\| \leq \rho(W) \|U-W\|.
\end{align*}
Now we can apply mean-value theorem to $\hbox{vec}\big(\triangledown C(W)\big)$. Let $W\in\mathbb{R}^{m \times n}$ and $w_{i}$ be the $i$-th entry of $\hbox{vec}\big(W\big)$, then there exist $Z_i$s such that $\hbox{vec}(Z_i)$s are on the line connecting $\hbox{vec}(W)$ and $\hbox{vec}(U)$ such that:
\begin{align*}
    \hbox{vec}\big(\triangledown C(U)\big) &= \hbox{vec}\big(\triangledown C(W)\big) + H(W)\hbox{vec}\big(U - W\big),
\end{align*}
with:
\begin{align*}
    &H(W)\\
    &= \begin{pmatrix} 
    \frac{\partial C}{\partial w_1 \partial w_1}(\hbox{vec}(Z_1))\ \ \ \ \cdots & \frac{\partial C}{\partial w_1 \partial w_{nm}}(\hbox{vec}(Z_1))
    \\
    \vdots & \vdots
    \\
    \frac{\partial C}{\partial w_{nm} \partial w_1}(\hbox{vec}(Z_{nm}))\ \cdots & \frac{\partial C}{\partial w_{nm} \partial w_{nm}}(\hbox{vec}(Z_{nm}))
    \end{pmatrix}.
\end{align*}
It is noted that $\|H(W)\| \leq \|\triangledown^2 C(W)\| + \rho(W)\delta(W)$ directly results from that all $\hbox{vec}(Z_i)$s lie in the ball $\mathcal{B}(W, \delta(W))$.
\end{proof}
\vspace{0.25cm}

Also The following helper lemma is required in the proof of Theorem~\ref{th:maml-convergence-stationary}.

\textbf{\begin{lemma}~\label{lem:Ci-F}
\textnormal{Let $W$ be task-stabilizing for all tasks $\mathcal{I}$ in Assumption~\ref{ass:tasks} and conditions in the Assumption~\ref{ass:variance} are met for that $W$. If $\eta < \min_{\mathcal{I}_i} \bigg\{ \dfrac{1}{2 \big(1 + \rho_{i}(W) \delta_{i}(W)\big) \|\triangledown^{2} C_i(W)\|} \bigg\}$, then we have:
\begin{align*}
    \mathbb{E}_{i}\big[\|\triangledown C_i(W)\|_{F}\big] 
    &< \dfrac{4 \big(\sigma + \| \triangledown F(W)\|_{F}\big)}{3}.
\end{align*}}
\end{lemma}}
\vspace{0.25cm}

\begin{proof}
We can write the followings:
\begin{align*}
    \| &\mathbb{E}_{i} \big[\triangledown C_i(W)\big] \|_{F}\\
    &\leq \| \mathbb{E}_{i} \big[ \triangledown F_{i}(W) \big] \|_{F} +  \| \mathbb{E}_{i} \big[ \triangledown C_i(W) - \triangledown F_{i}(W) \big] \|_{F} \\
    &= \| \triangledown F(W)\|_{F}\\
    &+  \| \mathbb{E}_{i} \big[ \triangledown C_i(W) - G_i(W) \triangledown C_{i}(W - \eta \triangledown C_{i} (W)) \big] \|_{F}\\
    &= \| \triangledown F(W)\|_{F}\\
    &+  \| \mathbb{E}_{i} \big[ \hbox{vec}\big(\triangledown C_i(W) - G_i(W) \triangledown C_{i}(W - \eta \triangledown C_{i} (W))\big) \big] \|\\ 
    &= \| \triangledown F(W)\|_{F}\\
    &+  \| \mathbb{E}_{i} \big[ \hbox{vec}\big(\triangledown C_i(W)\big)\\
    &\ \ \ \ \ \ \ \ \ \ - \big(I \otimes G_i(W)\big) \hbox{vec} \big(\triangledown C_{i}(W - \eta \triangledown C_{i} (W))\big) \big] \| \\
    &= \| \triangledown F(W)\|_{F}\\
    &+  \| \mathbb{E}_{i} \big[ \hbox{vec}\big(\triangledown C_i(W)\big)\\
    &\ \ \ \ \ \ \ \ \ \ - \big(I \otimes G_i(W)\big) \big(I - \eta H_i(W)\big)\hbox{vec} \big(\triangledown C_{i}(W)\big) \big] \| \\
    &= \| \triangledown F(W)\|_{F}\\
    &+  \| \mathbb{E}_{i} \big[\big(I - \big(I \otimes G_i(W)\big) \big(I - \eta H_i(W)\big)\big)  \hbox{vec} \big(\triangledown C_{i}(W)\big)\big] \| \\
    &\leq \| \triangledown F(W)\|_{F}\\
    &+   \mathbb{E}_{i} \big[ \|\big(I - \big(I \otimes G_i(W)\big) \big(I - \eta H_i(W)\big)\big)  \hbox{vec} \big(\triangledown C_{i}(W)\big) \| \big]\\
    &\leq \| \triangledown F(W)\|_{F} +   \gamma(W)\mathbb{E}_{i} \big[\| \triangledown C_{i}(W) \|_F \big],
\end{align*}
where $\gamma(W) = \max_{\mathcal{I}_i} \bigg\{\|I - \big(I \otimes G_i(W)\big) \big(I - \eta H_i(W)\big)\|\bigg\}$. Here we have used the fact that $\|\hbox{vec}(Z_1)\| = \|Z_1\|_{F}$, $\hbox{vec}(Z_1 Z_2) = (I \otimes Z_1) \hbox{vec}(Z_2)$ and applied Lemma~\ref{lem:mean-value-grad-cost} on the fourth equality. Also note that using the Assumption~\ref{ass:variance} we can conclude that $\mathbb{E}_{i}\big[\|\triangledown C_i(W)\|_{F}\big] \leq \sigma + \| \mathbb{E}_{i} \big[\triangledown C_i(W)\big] \|_{F}$ which together with $\| \mathbb{E}_{i} \big[\triangledown C_i(W)\big] \|_{F} \leq\| \triangledown F(W)\|_{F} +   \gamma(W)\mathbb{E}_{i} \big[\| \triangledown C_{i}(W) \|_F \big]$ results in:
\begin{align*}
    \| \mathbb{E}_{i} \big[\triangledown C_i(W)\big] \|_{F} \leq \dfrac{\| \triangledown F(W)\|_{F} + \sigma\gamma(W)}{1-\gamma(W)},
\end{align*}
We need $\gamma(W) < 1$. For this note that:
\begin{align*}
    \gamma(W) &= \max_{\mathcal{I}_i} \bigg\{\| \eta \big(H_{i}(W) + I \otimes \triangledown^{2} C_i(W)\big)\\
    &\ \ \ \ \ \ \ \ \ \ \ - \eta^{2} \big(I \otimes \triangledown^{2} C_i(W)\big) H_{i}(W) \| \bigg\}
\end{align*}
\begin{align*}
    &\leq \max_{\mathcal{I}_i} \bigg\{ \eta\big(\|H_{i}(W)\|+ \|\triangledown^{2} C_i(W)\| \big) \\
    &\ \ \ \ \ \ \ \ \ \ \ + \eta^{2} \|\triangledown^{2} C_i(W)\|\|H_{i}(W)\| \bigg\},
\end{align*}
and if $\eta$ satisfies the following condition:
\begin{align*}
    \eta &< \min_{\mathcal{I}_i} \bigg\{ \dfrac{1}{2 \big(\|H_{i}(W)\| + \|\triangledown^{2} C_i(W)\| \big)},\\
    &\ \ \ \ \ \ \ \ \ \ \ \ \dfrac{1}{2 \sqrt{\|H_{i}(W)\|\|\triangledown^{2} C_i(W)\|}} \bigg\}\\
    &= \min_{\mathcal{I}_i} \bigg\{ \dfrac{1}{2 \big(\|H_{i}(W)\| + \|\triangledown^{2} C_i(W)\| \big)} \bigg\}
\end{align*}
then $\gamma(W) < 3/4$. In addition note that according to Lemma~\ref{lem:mean-value-grad-cost} we have $\|H_{i}(W)\| \leq \|\triangledown^{2} C_i(W)\| + \rho_{i}(W) \delta_{i}(W)$, which concludes the proof.
\end{proof}

%% file: appendixC.tex
\section*{Appendix C}\label{app:pert-maml-grad-cost}

In this part we provide the proof for Lemmas~\ref{lem:beta} and ~\ref{lem:pert-maml-grad-cost}.

\begin{proof}(proof of Lemma~\ref{lem:beta})
Notice that the followings are equivalent:
\begin{itemize}
    \item $W'$ is stabilizing for pair $(A, B)$.
    \item $\hbox{rad}(A-BW') < 1$.
    \item There exists $P \succ 0$ such that 
    \begin{equation}\label{eq:lya}
        (A-BW')^{T} P (A-BW') - P \prec 0.
    \end{equation}    
\end{itemize}
Consider a candidate $P = P^{i}_{W} + \epsilon \bar{P}$ for some $\epsilon > 0$, where $P^{i}_{W}$ satisfies the Ricatti equation~(\ref{eq:ric}) for task $\mathcal{I}_{i}$ and parameter $W$ and $\bar{P} \succ 0$ satisfying:
\begin{align*}
    (A-BW)^{T} \bar{P} (A-BW) - \bar{P} = -I.
\end{align*}
Since $W$ is stabilizing for all tasks, such $\bar{P} \succ 0$ exists. Now let's look at the left hand side of ~(\ref{eq:lya}) with this candidate $P$. 
\begin{equation}\label{eq:X}
    \begin{aligned}
        (&A-BW')^{T} P (A-BW') - P\\ 
        &= (A-BW)^{T} P^{i}_{W} (A-BW) - P^{i}_{W}\\
        &+ \beta^{2} \triangledown F(W) ^{T} B^{T} P^{i}_{W} B \triangledown F(W)\\
        &+ \beta \triangledown F(W) ^{T} B^{T} P^{i}_{W} (A-BW)\\
        &+  \beta (A-BW)^{T}  P^{i}_{W} B \triangledown F(W)\\
        &+ \epsilon (A-BW)^{T} \bar{P} (A-BW) - \epsilon \bar{P}\\
        &+ \epsilon \bigg( \beta^{2} \triangledown F(W) ^{T} B^{T} \bar{P} B \triangledown F(W)\\
        &\ \ \ + \beta \triangledown F(W) ^{T} B^{T} \bar{P} (A-BW)\\
        &\ \ \ +  \beta (A-BW)^{T}  \bar{P} B \triangledown F(W)\bigg)\\
        &= - Q_{i} - W^{T}R_{i}W\\
        &+ \beta^{2} \triangledown F(W) ^{T} B^{T} P^{i}_{W} B \triangledown F(W)\\
        &+ \beta \triangledown F(W) ^{T} B^{T} P^{i}_{W} (A-BW)\\
        &+  \beta (A-BW)^{T}  P^{i}_{W} B \triangledown F(W)\\
        & - \epsilon I\\
        &+ \epsilon \bigg( \beta^{2} \triangledown F(W) ^{T} B^{T} \bar{P} B \triangledown F(W)\\
        &\ \ \ + \beta \triangledown F(W) ^{T} B^{T} \bar{P} (A-BW)\\
        &\ \ \ +  \beta (A-BW)^{T}  \bar{P} B \triangledown F(W)\bigg).    
    \end{aligned}
\end{equation}
With small enough $\beta$ such that:
\begin{align*}
    &\beta^{2} \triangledown F(W) ^{T} B^{T} P^{i}_{W} B \triangledown F(W)\\
    &+ \beta \triangledown F(W) ^{T} B^{T} P^{i}_{W} (A-BW)\\
    &+  \beta (A-BW)^{T}  P^{i}_{W} B \triangledown F(W) \prec - Q_{i} + W^{T}R_{i}W,
\end{align*}
the first four terms in ~(\ref{eq:X}) will be negative definite. Then with a small $\epsilon > 0$ the overall terms would be negative definite. Thus there exist $P \succ 0$ such that~(\ref{eq:lya}) is satisfied  for $W'$ and thus its is stabilizing for task $\mathcal{I}_{i}$. Not let's work on the condition for $\beta$: 
\begin{align*}
    \beta^{2} \|\triangledown F(W)\|\|B\| + 2\beta \|A-BW\| < \dfrac{\sigma_{min}(Q_{i})}{\|\triangledown F(W)\| \|B\| \| P^{i}_{W}\|}.
\end{align*}
A step-size $\beta$ satisfying
\begin{align*}
    \beta < \dfrac{\sqrt{\|A-BW\|^{2} + \frac{\sigma_{min}(Q_{i})}{P^{i}_{W}}} - \|A-BW\|}{\|\triangledown F(W)\| \|B\|},
\end{align*}
leads to an stabilizing $W'$. Also notice that $\|P^{i}_{W}\| \leq \frac{C_{i}(W)}{\mu}$ and this concludes the proof.
\end{proof}

\begin{proof}(proof of Lemma~\ref{lem:pert-maml-grad-cost})
Given that $W$ and $U$ are task-stabilizing for all tasks, the condition on step-size $\eta$ ensures that $W$ and $U$ are also MAML-stabilizing which is the direct result of Lemma~\ref{lem:suff-stable}. Now given that $\eta$ satisfy that bound we proceed further with the proof. We can conclude the followings by using $\triangledown F(W)$ definition in (\ref{eq:grad-F}) and rearranging the terms. 
\begin{equation*}
\begin{aligned}
    \|&\triangledown F(U)-\triangledown F(W)\|\\ 
                  &= \|\sum_{\mathcal{I}_i \sim p(\mathcal{I})} (\triangledown F_{i}(U) - \triangledown F_{i}(W))\|\\
                  &\leq \sum_{\mathcal{I}_i \sim p(\mathcal{I})} \|\triangledown F_{i}(U) - \triangledown F_{i}(W)\|\\
                  &\leq \sum_{\mathcal{I}_i \sim p(\mathcal{I})} \bigg(1 + \eta \|\triangledown^{2}C_{i}(U)\|\bigg)\\
                  &\ \ \ \ \ \ \ \times\|\triangledown C_{i}(U-\eta \triangledown C_{i}(U))- \triangledown C_{i}(W-\eta \triangledown C_{i}(W))\|\\
                  &\ \ \ \ \ \ \ +\eta \|\triangledown C_{i}(W-\eta \triangledown C_{i}(W))\|\|\triangledown^{2}C_{i}(U)-\triangledown^{2} C_{i}(W)\|.
\end{aligned}
\end{equation*}
We treat the terms separately. Let's start with $\|\triangledown^{2}C_{i}(U)-\triangledown^{2} C_{i}(W)\|$. According to the Assumption~\ref{ass:LQR-cost-assumptions}, we have:
\begin{align*}
    \|\triangledown^{2}C_{i}(U)-\triangledown^{2} C_{i}(W)\| \leq \rho_{i}(W) \|U-W\|,
\end{align*}
We can bound $\|\triangledown^{2}C_{i}(U)\| \leq \kappa_{i}(W)$
with $\kappa_{i}(W) =  \ell_{i}(W ) + \rho_{i}(W )\bar{\delta}_{i}(W )$. 
Now let's work on: 
 \begin{align*}
    \|\triangledown C_{i}&(W - \eta \triangledown C_{i}(W))\|\\ 
    &\leq \|\hbox{vec} \big(\triangledown C_{i}(W-\eta \triangledown C_{i}(W))\big)\|\\
    &= \|\hbox{vec} \big(\triangledown C_{i}(W)\big) - \eta H_{i}(W) \hbox{vec} \big(\triangledown C_{i}(W)\big)\|\\
    &\leq \|I - \eta H_{i}(W)\| \|\triangledown C_{i}(W)\|_{F},
\end{align*}
 where the first equality is based on Lemma~\ref{ass:LQR-cost-assumptions}. In addition, by applying the results of Lemma~\ref{ass:LQR-cost-assumptions} twice, we have:
\begin{align*}
    \|&\triangledown C_{i}(U - \eta \triangledown C_{i}(U)) - \triangledown C_{i}(W-\eta \triangledown C_{i}(W))\|\\ 
    &\leq \ell_{i}(W-\eta \triangledown C_{i}(W)) \|U - W - \eta (\triangledown C_{i}(U) - \triangledown C_{i}(W))\|\\
    &\leq \ell_{i}(W-\eta \triangledown C_{i}(W))(1 + \eta \ell_{i}(W)) \|U - W\|.
\end{align*}
Now combining these bound we have:
\begin{align*}
    \|\triangledown F(U)-\triangledown F(W)\| \leq L(W)\|U - W\|,
\end{align*}
where:
\begin{align*}
    L(W) &= \big(1 + \eta \kappa(W)\big)\bar{\ell}(W)(1+\eta \ell(W))\\
                  &\ \ \ +\eta \rho(W) (1 + \eta \|H(W)\|) \sum_{\mathcal{I}_i \sim p(\mathcal{I})} \|\triangledown C_{i}(W)\|_{F},
\end{align*}
with $\ell (W) = \max_{i} \ell_{i}(W)$, $\bar{\ell}(W) = \max_{i} \ell_{i}(W - \eta \triangledown C_{i}(W))$, $\kappa(W) = \max_{i} \kappa_{i}(W)$, $\|H(W)\| = \max_{i} \|H_{i}(W)\|$, and $\rho(W) = \max_{i} \rho_{i}(W)$.
\end{proof}

%% file: appendixD.tex
\section*{Appendix D}\label{app:maml-convergence-stationary}

In this part we provide the proof of Theorem~\ref{th:maml-convergence-stationary}.

\begin{proof}
Suppose $W_j$ is task-stabilizing for all LQR tasks, then the following condition $\beta_{j} < \min_{i} \bigg\{\dfrac{\alpha \delta_{i}(W_{j})}{\|\triangledown F(W_{j})\|}\bigg\}$
for some $\alpha \in (0, 1)$ is a sufficient condition on $\beta_{j}$ such that $W_{j+1}$ obtained after one step of gradient decent in ~(\ref{eq:gradient_step}) is also task-stabilizing for all LQR tasks. This is a direct result of Lemma~\ref{lem:suff-stable}. The second condition on $\eta_{j}$ captures the conditions on $\eta$ required for the Lemma~\ref{lem:pert-maml-grad-cost} and ensures that given a task-stabilizing $W_j$ for all LQR tasks, the $W_{j}$ and $W_{j+1}$ are also stabilizing for MAML objective. The remaining part of the proof shares a similar approach as in~\cite{fallah2020convergence}. As a result of Lemma~\ref{lem:pert-maml-grad-cost} and by applying Lemma 1.2.3 form~\cite{nesterov2003introductory}, we can write the following: 
\begin{align*}
    F(W_{j+1}) &\leq F(W_{j}) + \langle \triangledown F(W_{j}, W_{j+1} - W_{j}\rangle\\
    &+ \dfrac{L(W_j)}{2}\|U-W\|^2\\
               &= F(W_{j}) -\beta_{j} \langle \triangledown F(W_{j}), \triangledown F(W_{j})\rangle\\
               &+ \beta_{j}^{2}\dfrac{L(W_j)}{2} \langle \triangledown F(W_{j}), \triangledown F(W_{j})\rangle\\
               &= F(W_{j}) -\beta_{j} \bigg(1 - \beta_{j} \dfrac{L(W_j)}{2}\bigg) \| \triangledown F(W_{j})\|_{F}^{2}.
\end{align*}
If $\beta_{j}$ satisfy the conditions of the Assumption~\ref{ass:stepsize} for some $a$ and $b$ such that $2a > b^2$, we have:
\begin{align*}
    F(W_{j+1}) &\leq  F(W_{j}) -(2a - b^{2})\dfrac{ \| \triangledown F(W_{j})\|_{F}^{2}}{2L(W_{j})}.
\end{align*}
So far, we established that if the step-sizes meet certain conditions and the Algorithm~\ref{Al:MAML} starts with a task-stabilizing $W_0$ for all tasks, then the MAML objective will definitely decrease during the optimization process. Furthermore, if $W_0 \in \mathcal{W}$ then it is guaranteed that $W_j$s are also in $\mathcal{W}$ and thus we can utilize the Lemma~\ref{lem:Ci-F} to $\mathbb{E}_{i} \big[|\triangledown C_{i}(W_{j})|_{F}]$ to upper bound $L(W_j)$:
\begin{align*}
    2L(W_{j}) &< 2\bar{\ell}(W_{j})\big(1 + \eta_{j} \kappa(W_{j})\big)(1+\eta_{j} \ell(W_{j}))\\
    &\ \ +4 \eta_{j} \rho(W_{j})\big(\sigma + \| \triangledown F(W_{j})\|_{F}\big)\\
    &=: c_1(W_j, \eta_{j})\\
    &\ \ \ + c_2(W_j, \eta_{j})\big(\sigma +\| \triangledown F(W_{j})\|_{F}\big),
\end{align*}
where $c_1(W_j, \eta_{j}) = 2\bar{\ell}(W_{j})\big(1 + \eta_{j} \kappa(W_{j})\big)(1+\eta_{j} \ell(W_{j}))$ and $c_2(W_j, \eta_{j}) = 4 \eta_{j} \rho(W_{j})$.
Therefore we can continue as:
\begin{align*}
    &F(W_{j+1})\\ 
    &\leq  F(W_{j})\\
    &\ \ \ - \frac{ (2a - b^{2}) \| \triangledown F(W_{j})\|_{F}^{2}}{c_1(W_j, \eta_{j}) + + c_2(W_j, \eta_{j})\big(\sigma +\| \triangledown F(W_{j})\|_{F}\big)}\\
    &\leq  F(W_{j})\\
    &\ \ \ - \min \Bigg\{ \dfrac{ (2a - b^{2})\|\triangledown F(W_{j})\|_{F}^{2}}{c_{1}(W_j, \eta_{j}) + \sigma c_{2}(W_{j}, \eta_{j})},\\
    &\ \ \ \ \ \ \ \ \ \ \ \ \ \ \dfrac{ (2a - b^{2})\|\triangledown F(W_{j})\|_{F}}{c_{2}(W_{j}, \eta_{j})}\Bigg\}.
\end{align*}
Now if the condition in the theorem doesn't hold at iteration $j$, then we have $\|\triangledown F(W_{j})\|_{F} \geq \epsilon$, which implies that:
\begin{align*}
    &F(W_{j+1})\\ 
    &\leq  F(W_{j}) - \min \Bigg\{ \dfrac{(2a - b^{2})\epsilon^{2}}{c_{1}(W_j, \eta_{j}) + \sigma c_{2}(W_{j}, \eta_{j})} , \dfrac{ (2a - b^{2})\epsilon}{c_{2}(W_{j}, \eta_{j}) }\Bigg\}\\
    &\leq  F(W_{j}) -\dfrac{(2a - b^{2})\epsilon^{2}}{c_{1}(W_j, \eta_{j}) + c_{2}(W_{j}, \eta_{j}) \big(\sigma  + \epsilon\big)}.
\end{align*}
If we assume that for all iterations $0,...,T-1$ this condition does not hold then by summing both sides from $0$ to $T-1$ we obtain that:
\begin{align*}
    \sum_{j=0}^{T-1} F(W_{j+1}) 
    &\leq  \sum_{j=0}^{T-1} F(W_{j})\\ 
    &\ \ - \sum_{j=0}^{T-1}\dfrac{(2a - b^{2})\epsilon^{2}}{c_{1}(W_j, \eta_{j}) + c_{2}(W_{j}, \eta_{j}) \big(\sigma  + \epsilon\big)},
\end{align*}
which implies that
\begin{align*}
    \sum_{j=0}^{T-1}\dfrac{1}{c_{1}(W_j, \eta_{j}) + c_{2}(W_{j}, \eta_{j}) \big(\sigma  + \epsilon\big)} \leq \dfrac{F(W_{0}) - F(W_{T})}{(2a - b^{2})\epsilon^{2}}.
\end{align*}
We know that there exist $\tilde{c}_{1}$ and $\tilde{c}_{2}$ such that:
\begin{align}\label{eq:series}
    \sum_{j=0}^{T-1}\dfrac{1}{c_{1}(W_j, \eta_{j}) + c_{2}(W_{j}, \eta_{j}) \big(\sigma  + \epsilon\big)} = \dfrac{T}{\tilde{c}_{1} + \tilde{c}_{2} \big(\sigma  + \epsilon\big)},
\end{align}
which implies that:
\begin{align*}
    T \leq \dfrac{\big(F(W_{0}) - F(W_{T})\big)\big(\tilde{c}_{1} + \tilde{c}_{2} \big(\sigma  + \epsilon\big)\big)}{(2a - b^{2})\epsilon^{2}}.
\end{align*}
This means that if for all $j \in [0, T-1]$, the condition for $\|\triangledown F(W_{j})\|_{F}$ does not hold, then $T$ cannot exceed the right hand side of the above inequality. And this concludes the proof.
\end{proof}